\documentclass[english,aps,reprint,10pt,superscriptaddress,amssymb,amsfonts, twocolumn]{revtex4-1}
\usepackage[T1]{fontenc}
\usepackage[latin9]{inputenc}
\usepackage{mathrsfs}
\usepackage{mathtools}
\usepackage{amsmath}
\usepackage{amssymb}

\makeatletter
\usepackage[T1]{fontenc}
\usepackage[latin9]{inputenc}
\usepackage[normalem]{ulem}
\usepackage{textcomp}
\usepackage{graphicx,float}
\usepackage[dvipsnames,svgnames]{xcolor}
\usepackage{mathrsfs,mathtools,xfrac}
\usepackage{amsmath,amssymb,amsthm,amscd,bm,bbm}
\usepackage{txfonts} 
\usepackage{afterpackage,multirow,comment}
\usepackage{array,booktabs,minibox}
\newcolumntype{P}[1]{>{\centering\arraybackslash}p{#1}} 
\usepackage[colorlinks,bookmarks=false,citecolor=blue,linkcolor=blue,urlcolor=blue,filecolor=blue]{hyperref}
\usepackage{wasysym}

\usepackage {colortbl} 
\usepackage{lipsum}

\makeatother

\usepackage{babel}
\begin{document}
\title{
Anomalous quasiparticle lifetime in geometric quantum critical metals
}
\author{Hao Song}
\affiliation{
Department of Physics $\&$ Astronomy, McMaster University, Hamilton ON L8S 4M1, Canada} 
\affiliation{
CAS Key Laboratory of Theoretical Physics, Institute of Theoretical Physics, Chinese Academy of Sciences, Beijing 100190, China}
\author{Han Ma}
\affiliation{Perimeter Institute for Theoretical Physics, Waterloo ON N2L 2Y5, Canada}
\author{Catherine Kallin}
\affiliation{
Department of Physics $\&$ Astronomy, McMaster University, Hamilton ON L8S 4M1, Canada} 
\author{Sung-Sik Lee}
\affiliation{
Department of Physics $\&$ Astronomy, McMaster University, Hamilton ON L8S 4M1, Canada} 
\affiliation{Perimeter Institute for Theoretical Physics, Waterloo ON N2L 2Y5, Canada}
    
\vspace{5mm}

\date{\today}
\begin{abstract}
Metals can undergo geometric quantum phase transitions where the local curvature of the Fermi surface changes sign without a change in symmetry or topology. 
At the inflection points on the Fermi surface,
the local curvature vanishes, leading to 
an anomalous dynamics of quasiparticles.
In this paper, we study geometric quantum critical metals that support inflection points in two dimensions,
and show that the decay rate of quasiparticles 
goes as $E^{\alpha}$ with $1<\alpha<2$
as a function of quasiparticle energy $E$ 
at the inflection points.
\end{abstract}
\maketitle

In quantum critical metals,
critical fluctuations coupled with Fermi surfaces 
increase the incoherence of the single-particle excitations at low energies \cite{HOLSTEIN,HERTZ,
	PLEE1,
	REIZER,
	PLEE2,
    VARMALI,
	ALTSHULER,
	YBKIM,
	NAYAK,
	POLCHINSKI2,
 	MILLIS,
	ABANOV1,
	ABANOV3,
	ABANOV2,
	LOH,
	SENTHIL,
	DENNIS,
	MROSS,
	MAX0,
	MAX2,
	HARTNOLL,
	ABRAHAMS,
	JIANG,
	FITZPATRICK,
 	SSLEE,
	STRACK2,
	SHOUVIK2,
	PATEL2,
	SHOUVIK,
	RIDGWAY,
	HOLDER,
	PATEL,
	VARMA2,
	EBERLEIN,
	SCHATTNER,
	SHOUVIK3,
	CHOWDHURY,
	VARMA3,
 PhysRevLett.128.106402,
SENTHIL,
SUNGSIKREVIEW,
PhysRevX.11.021005,
2022arXiv220407585D}.
One example of a critical mode is 
the fluctuating order parameter 
 that becomes gapless 
 at a continuous 
 quantum phase transition 
 associated with a spontaneous symmetry breaking \cite{MILLIS,ABANOV1,ABANOV3,ABANOV2,LOH,SENTHIL,MROSS,MAX0,MAX2,HARTNOLL,DENNIS}.
Metals can also become critical as the topology of the Fermi surface changes 
without symmetry breaking.
Across topological phase transitions \cite{Tam2023topology,Tam2023probing},
the connectivity of the Fermi surface changes, generating van Hove singularities.
An enhanced low-energy density of states  gives rise to anomalous thermodynamic and transport behaviours as well as an enhanced superconductivity 
at topological critical points\cite{vanhove1953,Dzyaloshinski1987,schulz1987superconductivity,lederer1987antiferromagnetism,varma1989phenomenology,pattnaik1992evidence,gopalan1992effects,gonzalez1996renormalization,dzyaloshinskii1996extended,Furukawa1998,menashe1999fermi,Irkhin2002,kampf2003competing,le2009superconductivity,raghu2010superconductivity,nandkishore2012chiral,gonzalez2008kohn,yudin2014fermi,ghamari2015renormalization,kapustin2018wilsonian,barber2018,isobe2019supermetal,li2021high,jerzembeck2022superconductivity}.
In this paper, 
we consider 
a geometric quantum criticality associated with inflection points at which the local curvature of the Fermi surface vanishes.
We call metals with inflection points 
 {\it geometric quantum critical metals}.
They may arise as a stable phase without a fine turning,
and
a `trivial' metal without an inflection point and 
a geometrically critical phase  
 must be separated by a geometric quantum phase transition at which higher-order inflection points arise.
Such geometric phase transitions
connect different shapes of Fermi surfaces, 
for example, from a globally convex Fermi surface to  a peanut-shaped Fermi surface with locally concave segments  as is shown in Fig.~\ref{fig:FS}
\cite{fratini2002electronic,chubukov2006nonanalytic,roldan2006self}.
We show that quasiparticles at the inflection points remain coherent but they exhibit anomalously fast decay rates due to extra-soft particle-hole excitations 
present near the inflection points.

\begin{widetext}
\begin{center}
\begin{figure}[h]
\includegraphics[width=0.7\linewidth]{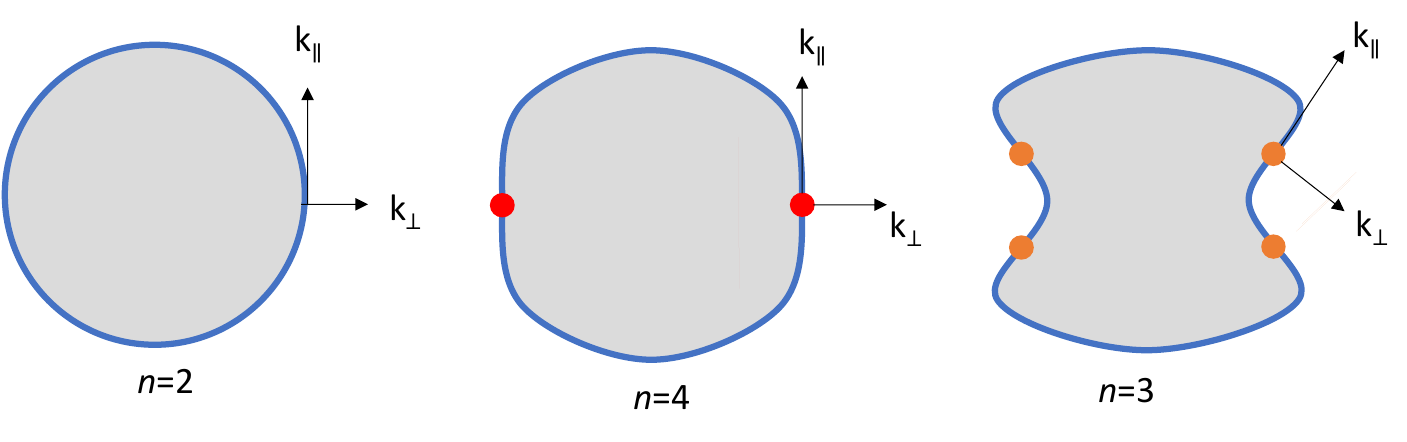}
	\caption{
A globally convex Fermi surface (left) undergoes a geometric quantum phase transition to a peanut-shaped Fermi surface (right) that supports four cubic inflection points denoted as dots.
At the critical point (middle), a pair of cubic inflection points merge into a quartic inflection point.
Near the $n$-th inflection point, 
the quasiparticle dispersion can be written as $E_{\bf K+\bf k} = k_\perp + k_\parallel^n$,
where 
${\bf K}$ is the Fermi momentum at the inflection point and
${\bf k}=(k_\perp,k_\parallel)$ denote the deviation of momentum away from the inflection point in the direction perpendicular and parallel to the Fermi surface, respectively.
}
\label{fig:FS}
\end{figure}
\end{center}
\end{widetext}

Let us consider a metal in two spatial dimensions described by 
\begin{equation}
H=\int_{\mathbf{p}}\xi_{\mathbf{p}}\psi_{\mathbf{p}}^{\dagger}\psi_{\mathbf{p}}+\int_{\mathbf{q},\mathbf{p}_{1},\mathbf{p}_{2}}\frac{V_{\mathbf{q}}}{2}\psi_{\mathbf{p}_{1}+\mathbf{q}}^{\dagger}\psi_{\mathbf{p}_{2}-\mathbf{q}}^{\dagger}\psi_{\mathbf{p}_{2}}\psi_{\mathbf{p}_{1}},
\end{equation}
where $\int_{\mathbf{p}}\equiv\int d^{2}p/\left(2\pi\right)^{2}$ etc. 
for brevity,
$V_{\mathbf{q}}=2\pi e^{2}/\left|\mathbf{q}\right|$ denotes
the bare Coulomb potential in the momentum space,  $\xi_{\mathbf{p}}$ is the bare
electron dispersion, and
$\psi_{\mathbf{p}}$ ($\psi_{\mathbf{p}}^{\dagger}$)
is the annihilation (creation)
operator of electron  
 with momentum
$\mathbf{p}$. 
Spin degrees of freedom are suppressed as it does not play an important role in our discussion.
Within the random-phase approximation valid in the weak coupling limit,
the self-consistent equations for the dressed propagator of electrons and the renormalized interaction can be written as
\begin{align}
G\left(\mathbf{p},\varepsilon\right) & =\frac{1}{-\varepsilon\left(1+i0^{+}\right)+\xi_{\mathbf{p}}+\Sigma\left(\mathbf{p},\varepsilon\right)},\label{eq:G}\\
\Sigma\left(\mathbf{p},\varepsilon\right) & =-i\int_{\mathbf{q},\omega}\mathcal{V}\left(\mathbf{q},\omega\right)G\left(\mathbf{p}-\mathbf{q},\varepsilon-\omega\right),\label{eq:self-energy}\\
\mathcal{V}\left(\mathbf{q},\omega\right) & =\frac{1}{V_{\mathbf{q}}^{-1}-\Pi\left(\mathbf{q},\omega\right)},\label{eq:V}\\
\Pi\left(\mathbf{q},\omega\right) & =-i\int_{\mathbf{p},\varepsilon}G\left(\mathbf{p},\varepsilon\right)G\left(\mathbf{p}+\mathbf{q},\varepsilon+\omega\right).\label{eq:Pi}
\end{align}
Here $G$, $\Sigma$, $\mathcal{V}$,
and $\Pi$ stand for the renormalized electron propagator, the self-energy, the dressed
interaction, and the polarization, respectively\cite{abrikosov2012methods}.

In Fermi liquids,
the dressed propagator can be written as
\begin{equation}
G\left(\mathbf{p},\varepsilon\right)\simeq\frac{Z_{\mathbf{p}}}{-\varepsilon+E_{\mathbf{p}}-i\,\mathrm{sgn}\left(\varepsilon\right)\tau_{\mathbf{p}}^{-1}},
\label{eq:G-1}
\end{equation}
where 
$E_{\mathbf{p}}$  is the renormalized energy dispersion
determined from 
$E_{\bf p}
=\xi_{\bf p} +
\textrm{Re}\Sigma ({\bf p},E_{\bf p})$
and 
$\tau_{\mathbf{p}}^{-1}=|\mathrm{Im}\,\Sigma(\mathbf{p},E_{\mathbf{p}}-i\,\mathrm{sgn}\left(\varepsilon\right)\tau_{\mathbf{p}}^{-1})|$
denotes the quasiparticle decay rate (which is equal to the reciprocal of quasiparticle lifetime). 
For Fermi liquids,
$\tau_{\mathbf{p}}^{-1}\ll|E_{\mathbf{p}}|$ 
in the limit that $\mathbf{p}$ approaches
the Fermi surface, which allows us to write the decay rate as
\begin{equation}
\tau_{\mathbf{p}}^{-1}\sim\left|\mathrm{Im}\,\Sigma\left(\mathbf{p},E_{\mathbf{p}}\right)\right|.
\label{eq:lifetime}
\end{equation}
For the globally convex Fermi surface, the decay rate is given by
\begin{equation}
\frac{1}{\tau_{\mathbf{p}}}\sim\begin{cases}
E_{\mathbf{p}}^{2}, & \text{in 3 or higher dimensions},\\
E_{\mathbf{p}}^{2}\ln \frac{1}{E_{\mathbf{p}}}, & \text{in 2 dimensions},
\end{cases}\label{eq:lifetimeFL}
\end{equation}
in the $E_{\mathbf{p}}\rightarrow0$ limit\cite{abrikosov2012methods,giuliani1982lifetime,fujimoto1990anomalous,zheng1996coulomb,menashe1996quasiparticle,narozhny2002interaction,galitski2004universal,chubukov2012first,li2013finite}.
In this paper, quasiparticle energies are measured relative to the Fermi surface.

Now we evaluate the decay rate near the inflection points that arise at and near the geometric quantum critical points.
Let ${\bf K}$ be the Fermi momentum at an inflection point of $n$-th order.
The dispersion of electrons near that point is written as
\begin{equation}
E_{\mathbf{K}+\mathbf{k}}=v_{F}k_{\perp}+a_{n}k_{\parallel}^{n},
\end{equation}
where ${\bf k}$ denotes the deviation of momentum away from the inflection point.
For simplicity, we have chosen a local cartesian coordinate system such that $k_\perp$ and $k_\parallel$ represents the momentum perpendicular and parallel to the Fermi surface (FS), respectively.
$v_{F}$ denotes
the Fermi velocity
and
$a_{n}k_{\parallel}^{n}$ captures the leading non-vanishing
dispersion along the tangential direction.
The curvature
of the $\mathrm{FS}$ at $\mathbf{K}$ is non-vanishing only in the presence
of term $a_{n}k_{\parallel}^{n}$ with $n=2$, and it is equal to $2\left|a_{2}\right|/v_{F}$.
For $n > 2$, the local curvature vanishes.

\begin{figure}
\includegraphics[width=1\columnwidth]{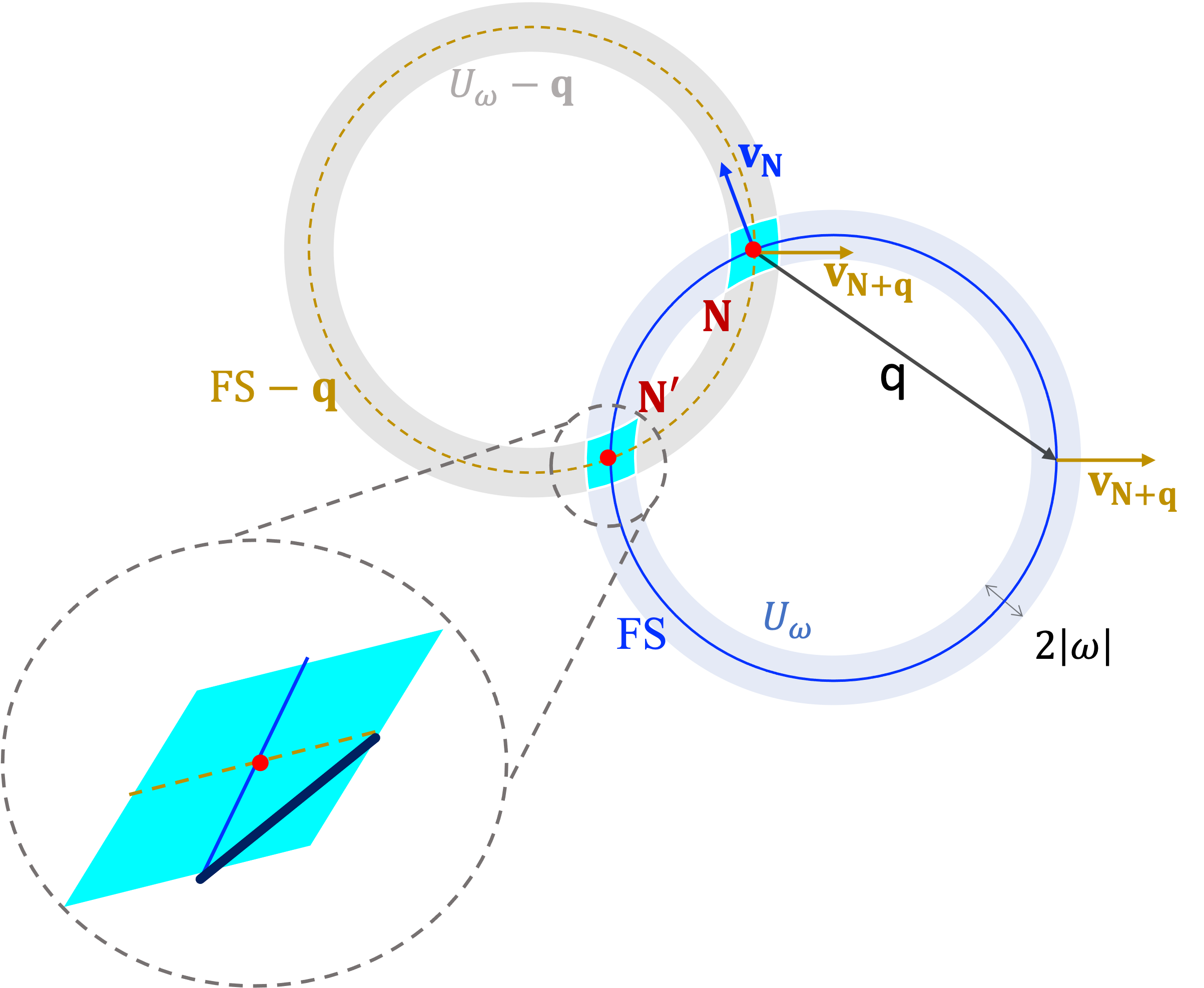}
\caption{
$\mathbf{N}$ and $\mathbf{N}'$ denote the intersection of 
the Fermi surface $\mathrm{FS}$ (solid blue circle) and
its translation $\mathrm{FS}-\mathbf{q}$ (dashed brown circle). 
The shells around the Fermi surfaces with thickness $2|\omega|$ create
diamonds centered around $\mathbf{N}$ and $\mathbf{N}'$.
The phase space for particle-hole pairs with momentum ${\bf q}$ 
 and energy $\omega$,
which is denoted as the thick line in the blow-up of the diamond,
is proportional to
$|\omega| / \left|\mathbf{v}_{\mathbf{N}'+\mathbf{q}}\times\mathbf{v}_{\mathbf{N}'}\right|$.
The smaller the angle between
$\mathbf{v}_{\mathbf{N}'+\mathbf{q}}$ and $\pm \mathbf{v}_{\mathbf{N}'}$,
the bigger the phase space becomes as the diamond gets more elongated.
}

\label{fig:Pi_integration}
\end{figure}

We begin by evaluating the polarization in Eq.~(\ref{eq:Pi})
to dress the interaction.
To the leading order in the interaction, we can approximate the renormalized electron propagator by 
$G\left(\mathbf{p},\varepsilon\right)\approx \frac{1}{-\varepsilon\left(1+i0^{+}\right)+E_{\mathbf{p}}}$
and express the polarization as 
\begin{equation}
\Pi\left(\mathbf{q},\omega\right)=\int_{\mathbf{p}}\frac{\Theta(-E_{\mathbf{p}+\mathbf{q}})-\Theta(-E_{\mathbf{p}})}{-\omega\left(1+i0^{+}\right)+E_{\mathbf{p}+\mathbf{q}}-E_{\mathbf{p}}},\label{eq:Pi-1}
\end{equation}
where $\Theta$ is the Heaviside step function. 
In the low frequency limit,
the real part of the polarization is given by the expression in the static limit ($\omega=0$), 
\begin{equation}
\Pi\left(\mathbf{q},0\right)=\int_{\mathbf{p}}\frac{\Theta(-E_{\mathbf{p}+\mathbf{q}})-\Theta(-E_{\mathbf{p}})}{E_{\mathbf{p}+\mathbf{q}}-E_{\mathbf{p}}}=-\int_{\mathbf{p}}\delta\left(E_{\mathbf{p}}\right),
\end{equation} 
which is the negative of the quasiparticle density of states at the Fermi surface for small $\mathbf{q}$.
This provides the screening of the Coulomb interaction.
The imaginary part is given by the on-shell particle-hole density of states.
By applying $\mathrm{Im}\frac{1}{x\pm i0^{+}}=\mp\pi\delta\left(x\right)$
to Eq.~(\ref{eq:Pi-1}),
we obtain 
\begin{gather}
\mathrm{Im}\,\Pi\left(\mathbf{q},\omega\right)=\pi\,\mathrm{sgn}\left(\omega\right)
\int_{\mathbf{p}}\left(\Theta(-E_{\mathbf{p}+\mathbf{q}})-\Theta(-E_{\mathbf{p}})\right)\delta\left(E_{\mathbf{p}+\mathbf{q}}-E_{\mathbf{p}}-\omega\right).\label{eq:ImPi}
\end{gather}
In order for the integrand to be nonzero, $E_{\mathbf{p}+\mathbf{q}}$
and $E_{\mathbf{p}}$ must have opposite signs. The delta function
further confines them to the interval $\left[-\left|\omega\right|,\left|\omega\right|\right]$. 
Thus, both $\mathbf{p}+\mathbf{q}$ and $\mathbf{p}$ lie within $U_{\omega}$,
where $U_{\omega}\coloneqq\{\mathbf{p}|\,|E_{\mathbf{p}}|\leq\left|\omega\right|\}$
is the narrow shell of the Fermi surface.
This requires $\mathbf{p}\in(U_{\omega}-\mathbf{q})\cap U_{\omega}$,
where $U_{\omega}-\mathbf{q}$ denotes the 
 set of momenta obtained by translating $U_{\omega}$
by $-\mathbf{q}$. 
Let us denote the set of 
momenta on the Fermi surface that is mapped into Fermi surface 
with the translation by $-{\bf q}$ as 
$
\mathscr{N}_{-\mathbf{q}}\coloneqq\left\{ \mathbf{N}\,|\,\mathbf{N}\in\left(\mathrm{FS}-\mathbf{q}\right)\cap\mathrm{FS}\right\} $.
For generic Fermi surfaces without perfect nesting,
$\mathscr{N}_{-\mathbf{q}}$ is a finite set.
Note $(U_{\omega}-\mathbf{q})\cap U_{\omega}\rightarrow \mathscr{N}_{-\mathbf{q}}$ as $\omega  \rightarrow 0$. Thus, 
for small $\omega$,
$(U_{\omega}-\mathbf{q})\cap U_{\omega}$ is made of 
disjoint regions, each of which is a neighborhood of one $\mathbf{N} \in \mathscr{N}_{-\mathbf{q}}$.
See Fig.~\ref{fig:Pi_integration} for an illustration. 
Accordingly, 
we can write the integration in Eq.~(\ref{eq:ImPi}) as
$\int_{\mathbf{p}}=\sum_{
 \mathbf{N} \in \mathscr{N}_{-\mathbf{q}} 
 }
 \int'_{\mathbf{p}}
 $, where 
 $\int'_{\mathbf{p}}$
represents the integration within the neighborhood of $\mathbf{N}$.
In this neighborhood,
we can use
$(E_{\mathbf{p}+\mathbf{q}},E_{\mathbf{p}})$ as independent variables for the two-dimensional momentum ${\bf p}$ to write
\begin{align}
& \mathrm{Im}\,\Pi\left(\mathbf{q},\omega\right) \nonumber
\\
= & \pi\,\mathrm{sgn}\left(\omega\right)
\sum_{ 
 \mathbf{N} \in \mathscr{N}_{-\mathbf{q}} 
 }
\int_{E_{\mathbf{p}+\mathbf{q}},E_{\mathbf{p}}}\frac{\Theta(-E_{\mathbf{p}+\mathbf{q}})-\Theta\left(-E_{\mathbf{p}}\right)}{\left|\mathbf{v}_{\mathbf{p}+\mathbf{q}}\times\mathbf{v}_{\mathbf{p}}\right|}\delta\left(E_{\mathbf{p}+\mathbf{q}}-E_{\mathbf{p}}-\omega\right)\nonumber \\
\sim &
\sum_{ 
 \mathbf{N} \in \mathscr{N}_{-\mathbf{q}} 
 }
 \frac{-\left|\omega\right|}{\left|\mathbf{v}_{\mathbf{N}+\mathbf{q}}\times\mathbf{v}_{\mathbf{N}}\right|}
\end{align}
with $\mathbf{v}_{\mathbf{p}}\coloneqq\partial_{\mathbf{p}}E_{\mathbf{p}}$,
the quasiparticle velocity. 
Here, $\left|\mathbf{v}_{\mathbf{p}+\mathbf{q}}\times\mathbf{v}_{\mathbf{p}}\right|$
represents the Jacobian and is approximated by its value at $\mathbf{N}$. 
Combining the real and imaginary parts,
we write
\begin{align}
\Pi\left(\mathbf{q},\omega\right) & \sim
-1-
i \sum_{
 \mathbf{N} \in \mathscr{N}_{-\mathbf{q}} }
\frac{\left|\omega\right|}{\left|\mathbf{v}_{\mathbf{N}+\mathbf{q}}\times\mathbf{v}_{\mathbf{N}}\right|},
\label{eq:Pi-2}
\end{align}
where the real part is set to be $-1$ as its value does not affect the scaling form of the decay rate that we are interested in.

The decay rate of the on-shell quasiparticle at momentum ${\bf p}$
can be written as
(see Appendix 
\ref{sec:self-energy}
for derivation)
\begin{equation}
\frac{1}{\tau_{\mathbf{p}}}\sim\int_{\mathbf{p}'}\left|\mathrm{Im}\,\mathcal{V}\left(\boldsymbol{\mathrm{p}}-\mathbf{p}',E_{\mathbf{p}}-E_{\mathbf{p}'}\right)\right|\Theta\left(E_{\mathbf{p}'}\left(E_{\mathbf{p}}-E_{\mathbf{p}'}\right)\right).
\label{eq:rlifetime}
\end{equation}
Here, 
$\mathbf{p}'$ 
and
$E_{\mathbf{p}'}$ 
represent the
momentum and energy of the virtual electron created when the electron at $\mathbf{p}$ 
emits the bosonic mode with momentum 
${\bf p}-{\bf p}'$
and energy
$E_{{\bf p}}-E_{{\bf p}'}$.
The decay rate is determined by 
$\mathrm{Im}\,\mathcal{V}(\boldsymbol{\mathrm{p}}-\mathbf{p}',E_{\mathbf{p}}-E_{\mathbf{p}'})$
that measures the
number of particle-hole excitations 
available for the scattering process.
Eq.~\eqref{eq:rlifetime} is convenient because the net decay rate is written as a sum of 
 non-negative contributions 
 from all intermediate states the external electron can 
 be scattered into.

\begin{center}
\begin{figure}[h]
\includegraphics[width=0.7\linewidth]{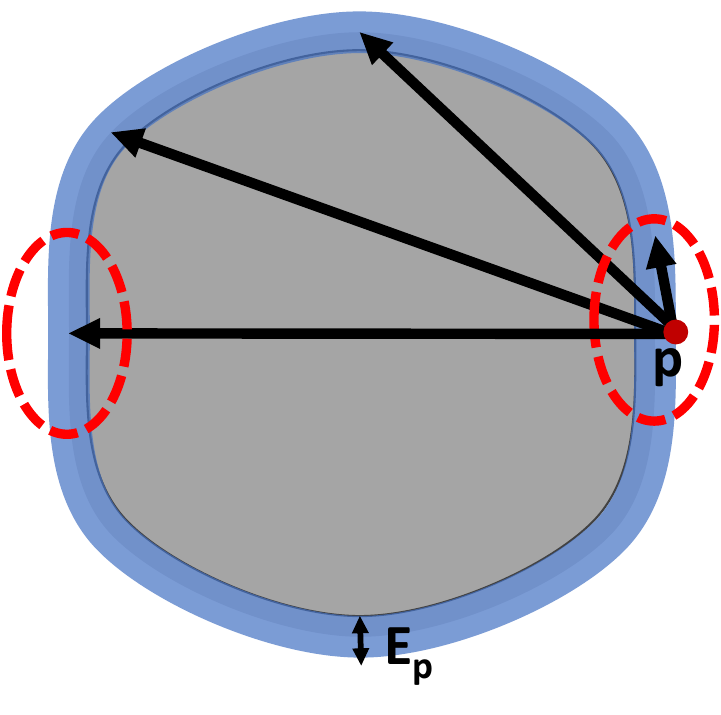}
	\caption{
A Fermi surface with the quartic inflection points.
The thin shell with thickness $E_{\bf p}$ around the Fermi surface represents the intermediate states that an electron with momentum ${\bf p}$ can be scattered into by creating a collective mode made of particle-hole excitations.
The shell can be divided into region $R^\parallel$ enclosed by dashed ellipsis  
and the remaining region $R^\nparallel$.
The dominant decay rate 
 of the quasiparticle at momentum ${\bf p}$ arises from the intermediate states in region $R^\parallel$ because of the large density of states of the particle-hole excitations available at low energies.
}
\label{fig:phase_space}
\end{figure}
\end{center}

We analyze the asymptotic behavior of $1/\tau_{\mathbf{p}}$ as $\mathbf{p}\rightarrow\mathbf{K}$ from above the Fermi surface,
where $\mathbf{K}$ denotes a Fermi momentum at an $n$-th inflection point.
At low energies,
one can approximate the
dressed interaction as
\begin{equation}
\mathcal{V}\left(\mathbf{q},\omega\right)\sim\frac{1}{\frac{\left|\mathbf{q}\right|}{2\pi e^{2}}+1-i\,\mathrm{Im}\Pi\left(\mathbf{q},\omega\right)}\sim\frac{1}{1-i\,\mathrm{Im}\,\Pi\left(\mathbf{q},\omega\right)}.
\label{eq:interaction}
\end{equation}
Note that the $\Theta$ function 
in Eq.~\eqref{eq:rlifetime} requires
$0 \leq E_{\mathbf{p}'} \leq E_{\mathbf{p}}$
for the integrand to be nonzero in Eq.~(\ref{eq:rlifetime}). Thus,
the $\mathbf{p}'$-integration is restricted to the thin shell of thickness $E_{\bf p}$ above the Fermi surface, which is denoted as $R$. 
Now, $R$ can be divided into 
two subsets as $R = R^{\parallel} \cup R^{\nparallel}$, where
$R^{\parallel}$ ($R^{\nparallel}$) denotes the collection of patches on the Fermi surface where 
${\bf p}-{\bf p}'$ 
connects
pairs of points on the Fermi surfaces that are parallel (not parallel).
This is illustrated in Fig.~\ref{fig:phase_space}.
The contribution from $R^\nparallel$ to the decay rate is at most
\begin{equation}
\frac{1}{\tau_{\mathbf{p}}^{\nparallel}} 
\sim
\int_{\mathbf{p}'\in R^\nparallel
}
\left|\mathrm{Im}\,\frac{\Theta\left(E_{\mathbf{p}'}\left(E_{\mathbf{p}}-E_{\mathbf{p}'}\right)\right)}{1-i\,\mathrm{Im}\,\Pi\left(\mathbf{p}-\mathbf{p}',E_{\mathbf{p}}-E_{\mathbf{p}'}\right)}\right|\sim E_{\mathbf{p}}^{2}.
\label{eq:R2}
\end{equation}
On the other hand,
the contribution from $R^\parallel$ is enhanced  
because $R^\parallel$ includes ${\bf p}'$ at which 
${\bf p}-{\bf p}'$ 
connects parallel or 
anti-parallel patches of Fermi surface such that
$ \left|\mathbf{v}_{\mathbf{p}}\times\mathbf{v}_{\mathbf{p}'}\right| =0$.
Let $\{ {\bf K }' \}$ be the set of momenta at which 
$ \left|\mathbf{v}_{\mathbf{K}}\times\mathbf{v}_{\mathbf{K}'}\right| =0$.
At least, $\{ {\bf K}'\} $ includes ${\bf K}$ and $-{\bf K}$.
The contribution 
from the patch centered at ${\bf K}'$ is controlled by the singularity of the polarization that goes as
\begin{equation}
\mathrm{Im}\,\Pi\left(\mathbf{p}-\mathbf{p}',\omega\right)\sim
-
\frac{\left|\omega\right|}{\left|\mathbf{v}_{\mathbf{K}}\times\mathbf{v}_{\mathbf{p}'}\right|}\sim
-
\frac{1}{a_m}
\frac{\left|\omega\right|}{\left|k_{\parallel}^{\prime m-1}\right|}.
\label{eq:18}
\end{equation}
Here, 
$k_{\parallel}^{\prime}$
denotes the component of ${\bf p}'-{\bf K}'$ that is tangential to the Fermi surface and
$m$ is the order of the inflection point in patch centered at ${\bf K}'$.
In the simple case 
in which 
only 
${\bf K}$
and 
$-{\bf K}$ 
are in the set of $\{ {\bf K}' \}$,
$m=n$.
In this case,
the contribution of $R^\parallel$ to the quasiparticle decay rate is given by
\begin{equation}
\frac{1}{\tau_{\mathbf{p}}^{\parallel}} 
\sim
\int_{E_{\mathbf{p}'},k_{\parallel}^{\prime}}\left|\mathrm{Im}\,\frac{\Theta\left(E_{\mathbf{p}'}\left(E_{\mathbf{p}}-E_{\mathbf{p}'}\right)\right)}{1-i\,
\frac{1}{a_n}
\frac{E_{\mathbf{p}}-E_{\mathbf{p}'}}{\left|k_{\parallel}^{\prime n-1}\right|}}\right|\sim\begin{cases}
\frac{1}{a_2}
E_{\mathbf{p}}^{2}\ln \frac{1}{E_{\mathbf{p}}}, & n=2,\\
\\
\frac{1}{a_n^{1/(n-1)}}
E_{\mathbf{p}}^{\frac{n}{n-1}}, & n>2
\end{cases}.
\label{eq:Rj}
\end{equation}
In the low-energy limit, 
the decay rate is dominated by
$1/\tau_{\mathbf{p}}^\parallel$.
For any finite $n$, quasiparticles remain well defined.
Nonetheless, for $n>2$, their decay rates are much bigger than those away from the inflection point,
\begin{equation}
\frac{1}{\tau_{\mathbf{p}}}\sim E_{\mathbf{p}}^{\frac{n}{n-1}}.
\label{eq:lifetimen}
\end{equation}
The enhanced decay 
 rate is due to the abundant low-energy particle-hole excitations available near the inflection points.
Within this analysis based on the approximate form of the polarization given in Eq. \eqref{eq:Pi-2}, 
the decay rate of quasiparticle 
 at momentum ${\bf p}$ is controlled by the inflection point of the highest order at which the Fermi velocity is parallel or anti-parallel to the Fermi velocity at ${\bf p}$.
For example, 
if the Fermi velocity at ${\bf p}$ is parallel to the Fermi velocities at 
a set of inflection points 
with order $m_i$,
the decay rate of the quasiparticle at ${\bf p}$
scales as $\frac{1}{\tau_{\mathbf{p}}}\sim E_{\mathbf{p}}^{\frac{m}{m-1}}$ where $m$ is the largest in $\{ m_i \}$.
If ${\bf p}$ itself is at the inflection point of order $n$, $m \geq n$.

Eq. \eqref{eq:lifetimen} can be checked through explicitly computation for simple cases.
Here, we consider a single patch theory with dispersion,
\begin{equation}
E_{\mathbf{K+k}}=k_{\perp}+k_{\parallel}^{n},
\end{equation}
where 
$\mathbf{k}$ 
denotes the deviation from an $n$-th inflection point,
$\mathbf{K}$.
For simplicity, we assume that 
$\mathbf{K}$ is the inflection point of the highest order.
In this case,
the polarization can be explicitly computed as
(see Appendix~\ref{sec:Polarization} for derivation)
\begin{equation}
\Pi\left(\mathbf{q},\omega\right)\sim-1- i \frac{\left|\omega\right|}{\left|q_{\parallel}^{n-1}\right|}\alpha_{n}\left(\frac{\omega-q_{\perp}}{q_{\parallel}^{n}}\right),
\end{equation}
where $\alpha_{n}$ is a real function defined by
\begin{align}
\alpha_{n}\left(x\right) & \coloneqq\begin{cases}
\frac{\Theta\left(x-2^{1-n}\right)}{\varrho_{n-1}\circ\varrho_{n}^{-1}\left(x\right)}, & n>1\text{ is odd},\\
\frac{1}{\varrho_{n-1}\circ\varrho_{n}^{-1}\left(x\right)}, & n>1\text{ is even}
\end{cases}
\end{align}
with 
\begin{equation}
\varrho_{n}\left(x\right)\coloneqq\left(x+1/2\right)^{n}-\left(x-1/2\right)^{n}.
\end{equation}
$\varrho_{n}^{-1}$ denotes the inverse function of $\varrho_{n}$.
For the first few $n$, they are given by $\varrho_{1}\left(x\right)=1$,
$\varrho_{2}\left(x\right)=2x$, $\varrho_{3}\left(x\right)=3x^{2}+\frac{1}{4}$,
and $\varrho_{4}\left(x\right)=3x^{3}+x$. 
When $n>1$ is even, the inverse function $\varrho_{n}^{-1}\left(x\right)$
is well-defined for all $x\in\mathbb{R}$. 
However, when $n>1$
is odd, $\varrho_{n}\left(x\right)$ is an even function bounded from
below by $\varrho_{n}\left(x\right)\geq2^{1-n}$. 
This leads to two possible values for the inverse function for $x\geq2^{1-n}$. 
Here, we choose the positive value for
$\varrho_{n}^{-1}\left(x\right)\geq0$.

For simplicity, let us choose $\mathbf{k}=\left(k_{\perp},0\right)$.
To compute $1/\tau_{\mathbf{p}}$ using Eqs.~(\ref{eq:rlifetime})
and (\ref{eq:interaction}), we first evaluate $\Pi\left(\mathbf{q},\omega\right)$
at $\mathbf{q}=\mathbf{k}-\mathbf{k}'$ and $\omega=E_{\mathbf{k}}-E_{\mathbf{k}'}$.
Noting $\omega=E_{\mathbf{k}}-E_{\mathbf{k}'}=\left(k_{\perp}-k_{\perp}^{\prime}\right)-k_{\parallel}^{\prime n}=q_{\perp}-k_{\parallel}^{\prime n}$,
we obtain 
$(\omega-q_{\perp})/q_{\parallel}^{n}=-k_{\parallel}^{\prime n}/(-k_{\parallel}^{\prime})^{n}=\left(-1\right)^{n-1}$
and $\alpha_{n}(\left(\omega-q_{\perp}\right)/q_{\parallel}^{n})=\alpha_{n}(\left(-1\right)^{n-1})$
is a nonzero constant independent on $\left(\mathbf{q},\omega\right)$.
Thus, we obtain
\begin{equation}
\Pi\left(\mathbf{q},\omega\right)\sim-1-\frac{i\left|\omega\right|}{\left|q_{\parallel}^{n-1}\right|}.
\end{equation}
Finally, the integration over the momentum of intermediate electron
in Eq.~(\ref{eq:Rj}) 
leads to 
Eq.~(\ref{eq:lifetimen}).

\begin{figure}[t]
\includegraphics[width=1\linewidth]{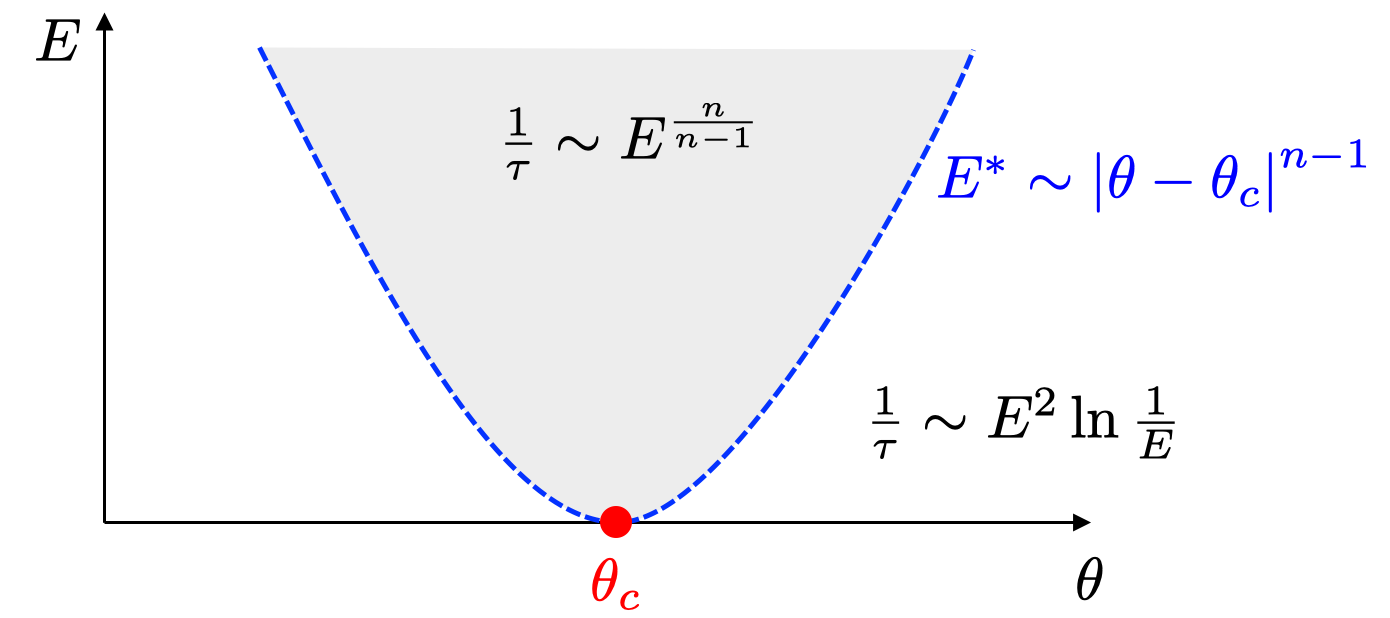}
	\caption{
A crossover of quasiparticle decay rate as a function of energy and angle around the Fermi surface.
Here, $\theta_c$ represents the angle for an $n$-th inflection point.
}
\label{fig:crossover}
\end{figure}

In summary, we 
 have demonstrated that quasiparticles at inflection points on the Fermi surface exhibit anomalously fast decay rates that scale as
$\frac{1}{\tau_{\mathbf{p}}}\sim E_{\mathbf{p}}^{\frac{n}{n-1}}$ with energy $E_{\bf p}$, 
where $n$ is the 
 order of the inflection point at ${\bf p}$.
Away from the inflection point, the Fermi surface acquires a small but non-zero curvature that is proportional to $a_2 \sim \delta \theta^{n-2}$ with $\delta \theta$ being the deviation of the angle away from the inflection point.
At a non-zero $\delta \theta$, the contribution from the quadratic part of the Fermi surface becomes dominant below a crossover energy scale $E^*$ determined from $E^{*2}/a_2 \sim E^{* n/(n-1)}$ up to a logarithmic correction.
This is expected to create a crossover in the decay rate as is shown in Fig.~\ref{fig:crossover}.

We conclude with a few remarks.
First, in the present work, it is assumed that the real part of the self-energy does not qualitatively modify the bare dispersion at the inflection point. 
This is consistent with the fact that the imaginary part of the self-energy is sub-leading compared to the bare term.
It is of interest to compute the leading non-analytic correction\cite{chubukov2003nonanalytic,chubukov2004singular} to the renormalized dispersion at the inflection points explicitly.
Second, one may also want to compute the decay rate of quasiparticles at inflection points 
for a closed Fermi surface beyond the patch approximation used in this work. 
Third, it would be 
 of great interest to see how the geometric quantum criticality manifests itself in collective modes 
 that describe fluctuations of Fermi surface shape.
Finally, our prediction can be tested by photoemission spectroscopy\cite{sobota2021angle}. 
Quasi-one-dimensional compounds with open Fermi surfaces generically possess cubic inflection points\cite{abrikosov1978conductivity,danner1994measuring,okazaki2004angle,j2006unconventional,Ishiguro1990}.
Quartic inflection points can in principle be created by driving a geometric phase transition 
 in layered materials through a uniaxial pressure\cite{barber2018,li2021high,jerzembeck2022superconductivity}.

\section*{Acknowledgement}

We acknowledge the support of the Natural Sciences  and Engineering Research Council of Canada. 
Research at the Perimeter Institute is supported in part by the
Government of Canada through Industry Canada, and by the Province of
Ontario through the Ministry of Research and Information. HS also acknowledges the support from the National Natural Science Foundation of China (Grant No.~12047503).

\bibliography{GQC}

\appendix
\onecolumngrid
\clearpage

\section{Electron self-energy 
\label{sec:self-energy}}

In this appendix, we derive 
Eq.~\eqref{eq:rlifetime}
for
the imaginary part of the electron self-energy. 
We use the spectral representation,
$f^{R}\left(\mathbf{p},\varepsilon\right)=\int\frac{d\varepsilon'}{\pi}\frac{\mathrm{Im}\,f^{R}\left(\mathbf{p},\varepsilon'\right)}{-\left(\varepsilon+i0^{+}\right)+\varepsilon'}$
for the retarded Green's functions,
where $f$ represents either the electron Green's function ($G$) 
or the propagator of the bosonic mode that mediates the interaction 
($\mathcal{V}$). 
Recall that $f\left(\mathbf{p},\varepsilon\right)=f^{R}\left(\mathbf{p},\varepsilon\right)\Theta\left(\varepsilon\right)+f^{R}\left(\mathbf{p},\varepsilon\right)^{*}\Theta\left(-\varepsilon\right)$,
where 
$f$ is the time-ordered Greeen's function and 
$\Theta$ is the Heaviside step function. 
This allows us to use
$f\left(\mathbf{p},\varepsilon\right)=\int\frac{d\varepsilon'}{\pi}\frac{\mathrm{Im}\,f^{R}\left(\mathbf{p},\varepsilon'\right)}{-\varepsilon\left(1+i0^{+}\right)+\varepsilon'}$
to express the self-energy in terms of the spectral functions as
\begin{align}
\Sigma\left(\mathbf{p},\varepsilon\right) 
 & =\int\frac{d^{2}q}{\left(2\pi\right)^{2}}\int\frac{d\omega'}{\pi}\int\frac{d\varepsilon'}{\pi}\frac{\mathrm{Im}\,\mathcal{V}^{R}\left(\mathbf{q},\omega'\right)\mathrm{Im}\,G^{R}\left(\mathbf{p}-\mathbf{q},\varepsilon'\right)}{-\varepsilon\left(1+i0^{+}\right)+\varepsilon'+\omega'}\left[\Theta\left(\varepsilon'\right)-\Theta\left(-\omega'\right)\right].
\end{align}
Applying the identity $\frac{1}{x\pm i0^{\dagger}}=\mathscr{P}\frac{1}{x}\mp i\pi\delta\left(x\right)$,
we obtain the imaginary part of the self-energy,
\begin{align}
\mathrm{Im}\,\Sigma\left(\mathbf{p},\varepsilon\right) 
 & =\int\frac{d^{2}q}{\left(2\pi\right)^{2}}\int\frac{d\omega'}{\pi}\,\mathrm{Im}\,\mathcal{V}^{R}\left(\mathbf{q},\omega'\right)\mathrm{Im}\,G^{R}\left(\mathbf{p}-\mathbf{q},\varepsilon-\omega'\right)\Theta\left(\left(\varepsilon-\omega'\right)\omega'\right),
\end{align}
where $\Theta\left(\varepsilon-\omega'\right)-\Theta\left(-\omega'\right)=\mathrm{sgn}\left(\varepsilon\right)\Theta\left(\left(\varepsilon-\omega'\right)\omega'\right)$
is used. 
In the low energy limit, 
we use
$\mathrm{Im}\,G^{R}\left(\mathbf{p}-\mathbf{q},\varepsilon-\omega'\right)
\approx 
\delta\left(\varepsilon-\omega'-E_{\mathbf{p}-\mathbf{q}}\right)$
to perform the frequency integration to obtain 
\begin{align}
\mathrm{Im}\,\Sigma\left(\mathbf{p},\varepsilon\right) & \approx \int\frac{d^{2}q}{\left(2\pi\right)^{2}}\,\mathrm{Im}\,\mathcal{V}^{R}\left(\mathbf{q},\varepsilon-E_{\mathbf{p}-\mathbf{q}}\right)\Theta\left(E_{\mathbf{p}-\mathbf{q}}\left(\varepsilon-E_{\mathbf{p}-\mathbf{q}}\right)\right).
\label{eq:A4}
\end{align}
Finally,
by noting
$\mathrm{sgn}\left(\mathrm{Im}\,\mathcal{V}^{R}\left(\boldsymbol{\mathrm{q}},\omega\right)\right)=-\mathrm{sgn}\left(\omega\right)$, we simplify
Eq. \eqref{eq:A4} as
\begin{align}
\mathrm{Im}\,\Sigma\left(\mathbf{p},\varepsilon\right) & \approx
-\mathrm{sgn}\left(\varepsilon\right)\int\frac{d^{2}p'}{\left(2\pi\right)^{2}}\,\left|\mathrm{Im}\,\mathcal{V}^{R}\left(\mathbf{p}-\mathbf{p}',\varepsilon-E_{\mathbf{p}'}\right)\right|\Theta\left(E_{\mathbf{p}'}\left(\varepsilon-E_{\mathbf{p}'}\right)\right).
\end{align}

\section{
Derivation of polarization in the patch theory\label{sec:Polarization}}

Consider a single patch with dispersion $E_{\mathbf{K}+\mathbf{k}}=k_{x}+k_{y}^{n}$,
where ${\bf K}$ is a mometum of an $n$-th inflection point 
and ${\bf k}$ is a deviation away from the inflection point.
The imaginary part of polarization $\mathrm{Im}\,\Pi\left(\mathbf{q},\omega\right)$
can be explicitly obtained 
from Eq.~(\ref{eq:ImPi}), 
\begin{align}
\mathrm{Im}\,\Pi\left(\mathbf{q},\omega\right) & = \pi\,\mathrm{sgn}\left(\omega\right)
\int_{E_{\mathbf{p}},k_{y}}\left(\Theta(-\omega-E_{\mathbf{p}})-\Theta(-E_{\mathbf{p}})\right)
\delta\left(q_{x}-\omega+\left(k_{y}+q_{y}\right)^{n}-k_{y}^{n}\right)\nonumber \\
 & \sim-\left|\omega\right|\int_{k_{y}}\delta\left(q_{x}-\omega+\left(k_{y}+q_{y}\right)^{n}-k_{y}^{n}\right)\stackrel{k_{y}\rightarrow q_{y}u}{=}-\frac{\left|\omega\right|}{\left|q_{y}^{n-1}\right|}\int_{u}\delta\left(\frac{q_{x}-\omega}{q_{y}^{n}}+\left(u+1\right)^{n}-u^{n}\right)\nonumber \\
 & \stackrel{u\rightarrow u-1/2}{=}-\frac{\left|\omega\right|}{\left|q_{y}^{n-1}\right|}\int_{u}\delta\left(\frac{q_{x}-\omega}{q_{y}^{n}}+\varrho_{n}\left(u\right)\right),
\end{align}
where $\mathbf{p}=\mathbf{K}+\mathbf{k}$ and $\varrho_{n}\left(u\right)$ denotes a function defined by
\begin{equation}
\varrho_{n}\left(x\right)\coloneqq\left(x+1/2\right)^{n}-\left(x-1/2\right)^{n}.
\end{equation}
Note that the range of $\varrho_{n}$ is $\mathbb{R}$ for even
$n>1$ and $[2^{1-n},\infty)$ for odd $n>1$.
Doing the delta function integration, we obtain
\begin{align}
\mathrm{Im}\,\Pi\left(\mathbf{q},\omega\right) & \sim-\frac{\left|\omega\right|}{\left|q_{y}^{n-1}\right|}\frac{1}{\left|\varrho_{n}^{\prime}\circ\varrho_{n}^{-1}\left(s_{n}\right)\right|}\times\begin{cases}
1, & n>1\text{ is even},\\
\Theta\left(s_{n}-2^{1-n}\right), & n>1\text{ is odd},
\end{cases}\\
 & \sim-\frac{\left|\omega\right|}{\left|q_{y}^{n-1}\right|}\frac{1}{\varrho_{n-1}\circ\varrho_{n}^{-1}\left(s_{n}\right)}\times\begin{cases}
1, & n>1\text{ is even},\\
\Theta\left(s_{n}-2^{1-n}\right), & n>1\text{ is odd},
\end{cases}
\end{align}
where $s_{n}\coloneqq\frac{\omega-q_{x}}{q_{y}^{n}}$ and we have
used $\varrho_{n}^{\prime}=n\varrho_{n-1}\sim\varrho_{n-1}$.

\end{document}